\documentclass[%
 reprint,
 showpacs,
 amsmath,amssymb,
prb,
]{revtex4-1}

\usepackage{dcolumn}
\usepackage[dvipdfmx]{graphicx}
\usepackage{mathrsfs}
\usepackage{accents}
\usepackage{bm}
\usepackage{color}

\makeatletter
\def\btt#1{\texttt{\@backslashchar#1}}
\DeclareRobustCommand\bblash{\btt{\@backslashchar}} \makeatother

\begin{document}

\title{Spontaneous edge current in a small chiral superconductor with a rough surface}

\author{Shu-Ichiro Suzuki$^1$ 
and Yasuhiro Asano$^{1,2,3}$}
\affiliation{$^1$Department of Applied Physics,
Hokkaido University, Sapporo 060-8628, Japan}
\affiliation{$^2$Center for Topological Science \& Technology,
Hokkaido University, Sapporo 060-8628, Japan}
\affiliation{$^3$Moscow Institute of Physics and Technology, 
141700 Dolgoprudny, Russia}
\date{\today}

\begin{abstract}
We study theoretically the spontaneous edge current in a small chiral superconductor with
surface roughness.  We obtained self-consistent solutions of the pair potential and the vector potential by
solving the quasiclassical Eilenberger equation and the Maxwell equation simultaneously. We
then employed them to calculate numerically the spatial distribution of the chiral edge current
in a small superconductor.  The characteristic behavior of the spontaneous edge current depends
strongly on the symmetries of the order parameters such as chiral $p$\,-, chiral $d$\,- and
chiral $f$-wave pairing.  The edge current is robust under the surface roughness in the chiral
$p$\,- and chiral $d$\,-wave superconductors.  In the chiral $d$\,-wave case, the surface
roughness tends to flip the direction of the chiral current.  On the other hand, the edge
current in a chiral $f$-wave superconductor is fragile when there is surface roughness.  We
also discuss the temperature dependence of a spontaneous magnetization, which is a
measurable value in standard experiments. 
\end{abstract}


\maketitle

\section{Introduction}
The experimental detection of a spontaneous edge current could be direct evidence of chiral
superconductivity.  A number of Cooper pairs sharing a specific angular momentum carry the
spontaneous edge current in chiral superconductors
\cite{M.Matsumoto_JPSJ_1999,A.Furusaki_PRB_2001} 
which can be experimentally measured as spontaneous magnetization.  Strontium ruthenate
Sr$_2$RuO$_4$ is a leading candidate for a chiral $p$\,-wave superconductor
\cite{Y.Maeno_Nature_1994,T.M.Rice_JP_1995,A.P.Mackenzie_RevModPhys_2003} 
whose pair potential is described by $\Delta(k_x \pm i k_y) = \Delta e^{i \chi \theta}$ in
momentum space.  Here $k_x = \cos \theta$ ($k_y = \sin \theta$) is the normalized wavenumber in
the $x$ ($y$) direction, and $\Delta$ is the amplitude of the pair potential. The topological
Chern number $\chi=1$ or $-1$ corresponds to the angular momentum of a Cooper pair.  In
addition to chiral $p$\,-wave superconductivity, the possibilities of chiral $d$\,-wave
($\chi=\pm 2$) and chiral $f$-wave ($\chi= \pm 3$) superconductivity have been discussed in
recent experiments.  
\cite{K.Takeda_Nature_2003,G.Baskaran_PRL_2003,M.Ogata_JPSJ_2003,A.Tanaka_PRL_2003,R.Nandkishore_NarPhys_2012,M.L.Kiesel_PRB_2012,M.L.Kiesel_PRL_2013,M.H.Fischer_2014_PRB,Y.Kasahara_NJP_2009,H.Kusunose_JPSJ_2012,M.Tsuchiizu_PRB_2015}
Several theories have suggested that the amount of 
edge current become smaller in a chiral superconductor with a larger $|\chi|$.
\cite{Y.Tada_PRL_2015,W.Huang_PRB_2014} 
However, unfortunately, no spontaneous chiral current has yet been experimentally observed. 
\cite{P.G.Bjornsson_PRB_2005,J.R.Kirtley_PRB_2007}

The absence of spontaneous magnetization in experiments has mainly been attributed to three
effects: (i) the Meissner screening of the edge current by the bulk superconducting condensate,
(ii) the reduction of the chiral current by the potential disorder near the surface of a
superconductor, and (iii) the complicated electronic structures of superconductors.
 The first effect was partially studied by Matsumoto and Sigrist.
\cite{M.Matsumoto_JPSJ_1999}
They theoretically confirmed a reduction in the edge current caused by the Meissner effect in a
chiral $p$\,-wave superconductor. However, the resulting spontaneous magnetization is large
enough to be measured in experiments.  The second effect is linked to the issue of the
intrinsic angular momentum in the $^3$He-A phase. \cite{J.A.Sauls_PRB_2011,K.Nagai_JLTP_2014}
Experimentally it is difficult to make a superconducting sample with a specular surface. For
instance, a small ruthenate superconductor cluster can be fabricated by using the focused ion
beam technique, \cite{K.Saitoh_APE_2012,K.Saitoh_PRB_2015} which would seriously damage the
sample quality near the surface. Several theoretical papers have already suggested the
presence of edge states in a chiral $p$\,-wave superconductor when there is surface
roughness. \cite{Ashby_PRB_2009,S.V.Bakurskiy_PRB_2014} On the other hand, 
when a chiral $p$\,-wave superconductor is covered by a clean normal metal, 
the chiral current is dramatically reduced.\cite{S.Lederer_PRB_2014} The third effect 
has been discussed specifically in Sr$_2$RuO$_4$. It has been known that the gap 
anisotropy\cite{A.Bouhon_PRB_2014,W.Huang_PRB_2015} and the multiband 
structures~\cite{T.Scaffidi_PRL_2015} suppress the chiral edge current. 
Even today, we do not know how the
Meissner screening and the surface roughness reduce the edge current in chiral $d$\,- and
$f$-wave superconductors. In previous papers, \cite{S.I.Suzuki_PRB_2014,S.I.Suzuki_PRB_2015}
we studied the Andreev bound states 
\cite{L.J.Buchholtz_PRB_1981,J.Hara_PTP_1986,S.Kashiwaya_PCS_1998,C.R.Hu_PRL_1994,Y.Tanaka_PRL_1995,Y.Asano_PRB_2004}
(ABSs) in time-reversal non-chiral superconductors characterized by $d_{x^2-y^2}$-wave or
$p_x$-wave pair potentials.  We found that the ABSs in a $p_x$-wave superconductor are robust
even in the presence of surface roughness, whereas those in a $d$\,-wave superconductor are fragile
against surface roughness.  This conclusion is well explained by the symmetry of the Cooper
pairs induced near the surface.  However, it is unclear if it is possible to generalize our
conclusions straightforwardly to chiral superconductors.  We will address these issues in the
present paper.

In this paper, we theoretically study the spontaneous edge currents and the spontaneous
magnetization in a small chiral superconducting disk based on the quasiclassical Eilenberger
formalism. To discuss the relation between the pairing symmetry and the sensitivity of the
chiral edge current to the surface roughness, we consider the simple chiral order parameters on
a circular shaped Fermi surface.  By solving the Eilenberger and Maxwell equations
self-consistently and simultaneously, we obtain the spatial profiles of the chiral edge
currents and the temperature dependence of a spontaneous magnetization. The surface roughness
is considered in terms of the impurity self-energy of a quasiparticle.  To define the
magnetization of a sample, we need to consider a finite-size superconductor such as disks.
Moreover, setting the radius of a disk to be comparable to the coherence length allows us to
justify the assumption that there is no chiral-domain wall in a disk. 
We conclude that the robustness of the spontaneous edge current depends strongly on the paring
symmetry. In a chiral $p$\,-wave superconductor, the amplitude of the chiral current in a disk
with a rough surface is comparable to that in a disk with a specular surface. In a chiral
$d$\,-wave superconductor, there are two edge channels in a disk with a specular surface.  They
carry the chiral currents in opposite directions.  In the presence of surface roughness, one
channel near the surface disappears and the other channel far from the surface carries the
robust chiral current.  We show that the surface roughness changes the net-current direction in
a chiral $d$\,-wave disk. The edge current in a chiral $f$-wave superconductor is fragile in
the presence of surface roughness.  The effects of Meissner screening on the chiral edge
current depend on the spatial current distribution near the surface.  When the current
decreases monotonically with increases in the distance from the surface, the Meissner effect
always reduces the chiral current. On the other hand, when the chiral current changes its
direction as a function of distance from the surface, the Meissner screening effect becomes
weaker.  Such a complicated current distribution causes the self-screening effect among edge
currents flowing in opposite directions.

This paper is organized as follows. 
In Sec.~II, we explain the quasiclassical Eilenberger formalism and define the spontaneous 
magnetization of a small superconducting disk. 
In Sec.~III, we present results obtained using non-self-consistent simulations 
(i.e., with a homogeneous pair potential and without a vector potential). 
In Sec.~IV, we discuss the spontaneous edge current in a superconducting disk with a 
{\it specular} surface. 
In Sec.~V, we study the effects of surface roughness on the spontaneous edge current. 
In Sec.~VI, we demonstrate the temperature dependence of a spontaneous magnetization, 
which is a measurable value in experiments. 
In Sec.~VII, we summarize this paper.

\section{Quasiclassical Eilenberger theory}
\begin{figure}[t]
	\begin{center}
	\includegraphics[width=0.45\textwidth]{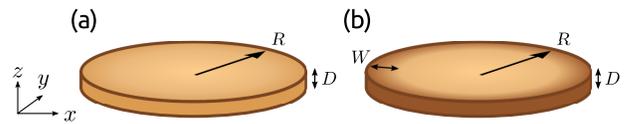}
		\caption{
      Schematics of superconducting disks.  The disk with a specular surface and that with a
      rough surface are shown in (a) and (b), respectively.  The radius and thickness of a disk
    are denoted by $R$ and $D$, respectively. The width of the disordered region is denoted
    by $W$ in (b).  The small enough radius allows us to assume that there is no chiral domain
  in the disk. }
	\label{fig:sche}
  \end{center}
\end{figure}

Let us consider a small chiral superconducting disk as shown in Fig.~\ref{fig:sche}.  We assume
that there is no chiral domains by choosing the radius of the disk $R$ to be comparable to the
coherence length $\xi_0=\hbar v_F/2\pi T_c$, where $v_F$ is the Fermi velocity and $T_c$ is the
superconducting transition temperature.  We apply the quasiclassical Green function theory of
superconductivity \cite{G.Eilenberger_1968} to calculate the edge current of a chiral
superconductor. 
In an equilibrium superconductor, the Eilenberger equation takes the form
\begin{align}
	i v_F\boldsymbol{k} \cdot \hspace{-1mm}
  \boldsymbol{\nabla}_{\boldsymbol{r}} \, \check{{g}}
  +\left[ \check{H} + \check{\Sigma}\, , \check{g} \, \right]_- =0,
  \label{eilenberger_eq}
\end{align}
where $v_F$ is the Fermi velocity, $\boldsymbol{k}$ is the unit wave vector on the Fermi
surface, and $[\alpha, \beta]_- = \alpha \beta - \beta \alpha$.  We employ the isotropic
cylindrical Fermi surface	(i.e., no $k_z$ dependence) as studied in
Ref.~\onlinecite{M.Matsumoto_JPSJ_1999}
{ because most chiral superconductors are layered materials. }
Throughout this paper, we use the set of units $\hbar = k_B = c = 1$, where $ 2\pi \hbar $ 
is the Planck constant, $k_B$ is the Boltzmann constant, and $c$ is the speed of light. 
The matrices $\check{g}$ and $\check{H}$ are defined as follows, 
\begin{align}
  \check{g}(\boldsymbol{r},{\boldsymbol{k}},i\omega_n)
  &=\left[\begin{array}{cc}
                 \hat{g} (\boldsymbol{r},{\boldsymbol{k}},i\omega_n) & 
                 \hat{f} (\boldsymbol{r},{\boldsymbol{k}},i\omega_n) \\
    -\undertilde{\hat{f}}(\boldsymbol{r},{\boldsymbol{k}},i\omega_n) & 
    -\undertilde{\hat{g}}(\boldsymbol{r},{\boldsymbol{k}},i\omega_n)
  \end{array}\right],
  \label{g_def} \\[2mm]
  \check{H}(\boldsymbol{r},{\boldsymbol{k}},i\omega_n)
  &=\left[\begin{array}{cc}
                \hat{\xi   } (\boldsymbol{r},{\boldsymbol{k}},i\omega_n) & 
                \hat{\Delta} (\boldsymbol{r},{\boldsymbol{k}})           \\
    \undertilde{\hat{\Delta}}(\boldsymbol{r},{\boldsymbol{k}})           & 
    \undertilde{\hat{\xi   }}(\boldsymbol{r},{\boldsymbol{k}},i\omega_n)
  \end{array}\right],
  \label{H_def}
\end{align}
with $\hat{\xi}(\boldsymbol{r},{\boldsymbol{k}},i\omega_n) = \left[ i\omega_n + ev_F
\boldsymbol{k} \cdot \boldsymbol{A}(\boldsymbol{r}) \right] \hat{\sigma}_0$, where
$\omega_n=(2n+1)\pi T$ is the Matsubara frequencies with $n$ being an integer, $T$ is the
temperature, $\hat{\Delta}$ represents the pair potential, $\hat{\sigma}_0$ is the $2\times 2$
identity matrix in spin space, and $\boldsymbol{A}$ is the vector potential induced by the
chiral edge current.  We introduce the definition
$\undertilde{K}(\boldsymbol{r},{\boldsymbol{k}},i\omega_n)
=K^\ast(\boldsymbol{r},-{\boldsymbol{k}},i\omega_n)$.  
The symbol $\check{\cdot}$ represents a $ 4 \times 4$ matrix structure in particle-hole space
and the symbol $\hat{\cdot}$ represents a $2 \times 2$ matrix structure in spin space.

We consider three
chiral superconductors with different pairing symmetries: spin-triplet chiral $p$\,-wave,
spin-singlet chiral $d$\,-wave, and spin-triplet chiral $f$-wave pairings.  In the spin-triplet
superconductor, we assume that the pairing interactions work between two electrons with
opposite spins.  This assumption does not loose any generality of the argument below.  The pair
potential are described by
\begin{align}
  \hat{\Delta}(\boldsymbol{r},\theta) = \left\{ 
	\begin{array}{cl}
		\Delta(\boldsymbol{r},\theta)  \hat{\sigma}_1 & \text{ for a spin triplet, } \\[2mm]
		\Delta(\boldsymbol{r},\theta) i\hat{\sigma}_2 & \text{ for a spin singlet, } \\
	\end{array}\right.
	\label{eq:spin}
\end{align}
where $ \hat{\sigma}_j$ for $j = 1$-$3$ are the Pauli matrices in spin space.  The matrix Green
functions in Eq.~(\ref{g_def}) can be represented by the scaler Green functions as 
\begin{align}
  \hat{g}(\boldsymbol{r},\theta, i\omega_n) 
	& = {g}(\boldsymbol{r},\theta, i\omega_n) \hat{\sigma}_0, \\
  \hat{f}(\boldsymbol{r},\theta, i\omega_n) 
	& = \left\{ 
	\begin{array}{cl}
		f(\boldsymbol{r},\theta, i\omega_n) (-i\hat{\sigma}_1) & \text{for a triplet } \\[2mm]
		f(\boldsymbol{r},\theta, i\omega_n)    \hat{\sigma}_2  & \text{for a singlet. } \\
	\end{array}\right.
  \label{eq:spin_gf}
\end{align}
The pair potential in a chiral superconductor is described by 
\begin{align}
  {\Delta}(\boldsymbol{r},\theta) = 
	    	\Delta_{1}(\boldsymbol{r}) \cos( \chi \theta) 
		+ i \Delta_{2}(\boldsymbol{r}) \sin( \chi \theta),  
		\label{eq:pp_k}
\end{align}
where $\theta$ is the azimuthal angle in the momentum space (i.e.,
$k_x = \cos \theta$ and $k_y = \sin \theta$), and $\Delta_{1}$ and $\Delta_{2}$ are the local
amplitudes of two independent components.  The
topological numbers $\chi = \pm 1$, $\pm 2$, and $\pm 3$ characterize the chiral $p$\,-, chiral
$d$\,-, and chiral $f$\,-wave superconductivity, respectively.  The doubly degenerate chiral
superconducting states are indicated by $\pm \chi$.  In this study, we consider superconducting
states with a positive $\chi$.
Deep inside a superconductor (i.e., bulk region), the relation $\Delta_{1} = \Delta_{2}$ is satisfied. 
Therefore the pair potentials in the bulk are represented as 
\begin{align}
  {\Delta}(\theta) 
	= \bar{\Delta}(T) e^{i \chi \theta}, 
\end{align}
where $\bar{\Delta}(T)$ is the amplitude of the uniform pair potential at a temperature $T$.
{The amplitude of the superconducting gap is isotropic in momentum space.}
In the simulations, $\Delta_1$ and $\Delta_2$ are self-consistently determined by the gap equation, 
\begin{align}
  \left[ \begin{array}{c}
		\Delta_1( \boldsymbol{r} ) \\[1mm]
	  \Delta_2( \boldsymbol{r} )
  \end{array} \right]
  \hspace{-1mm}=\hspace{-1mm} N_0 g_0 \pi T \sum_{\omega_n} 
  \int \hspace{-1mm}\frac{d\theta'}{2 \pi}
  f( \boldsymbol{r}, \theta', i\omega_n)\hspace{-1mm}
  \left[ \begin{array}{c}
		V_1( \theta' ) \\[1mm]
	  V_2( \theta' )
  \end{array} \right]
	\label{gap}
\end{align}
where $N_0$ is the density of states per spin at the Fermi level in three-dimension. 
The coupling constant $g_0$ is determined by
\begin{align}
	(N_0 g_0)^{-1}
	=  
  \mathrm{ln}
  \left( \frac{T}{T_c} \right)
  +\sum_{n = 0}^{n_c}
	\frac{1}{n+1/2}, 
	\label{couple}
\end{align}
where $ n_c = ( \omega_c / 2 \pi T )$ 
with $\omega_c$ being the cutoff energy. 
The functions $V_1$ and $V_2$ represent the attractive interactions as
\begin{align}
  V_1(\theta) = 2\cos( \chi \theta ), \quad
  V_2(\theta) = 2\sin( \chi \theta).
\end{align}
{
In our model, the Andreev bound states never appear at the surface in the $z$ direction (i.e., the top
and bottom surfaces in Fig.~\ref{fig:sche}) because the pair potential
does not depend on $k_z$.\cite{Y.Asano_PRB_2004} 
As a result, the pair potential is less dependent on $z$. Thus, by setting the disk thin
enough $ D \lesssim \xi_0 < \lambda_L$, we ignore the $z$ dependence of the quasiclassical Green
functions. 
However the thickness of
the disk need to be larger than the Fermi wavelength $1/k_F$, so that the quasiclassical theory
can be applied.  \cite{A.F.Andreev_JETP_1964,Y.Nagato_PRB_1995}
}

The effects of the rough surface are taken into account through the impurity self-energy, which
is defined by
\begin{align}
	\check{\Sigma} (\boldsymbol{r}, i\omega_n)
	= \left\{ \begin{array}{cl}
      \displaystyle \frac{i}{2 \tau_0}
      \displaystyle \int \cfrac{d \theta}{2\pi} 
			\, \check{g} (\boldsymbol{r}, \theta, i\omega_n ) & 
			\text{ for $ r > R-W$, } \\[5mm]
			0 & 
			\text{ for $ r < R-W$, }
		\end{array} \right.
\end{align}
where $r= (x^2 + y^2)^{1/2}$ and $\tau_0$ is the mean free time due to the
impurity scatterings.  The self-energy has finite values only near the surface as shown in
Fig.~\ref{fig:sche}(b), where $W$ is the width of the disordered region.

The electric current $\boldsymbol{j}(\boldsymbol{r})$ is calculated from the Green function
\begin{align}
 \bm{j}(\bm{r})=
 \frac{\pi e v_F N_0}{2i} 
 T \sum_{\omega_n} 
 \int \hspace{-1mm}\frac{ d \theta }{2 \pi }
 \textrm{Tr} \hspace{-1.0mm}\left[ 
   \check{T}_3\, \bm{k}\, \check{g}( \bm{r}, \theta, i\omega_n )
 \right],
 \label{eq:j}
\end{align}
where $\check{T}_3 = \text{diag}[\hat{\sigma}_0,-\hat{\sigma}_0]$. 
The vector potential is determined by solving the Maxwell equation,
\begin{align}
  \nabla \times \boldsymbol{A}(\boldsymbol{r}) =&~      \boldsymbol{H}(\boldsymbol{r}),\\
  \nabla \times \boldsymbol{H}(\boldsymbol{r}) =&~ 4\pi \boldsymbol{j}(\boldsymbol{r}). 
	\label{mag_j}
\end{align}
In a finite size superconductor, we define the
amplitude of a spontaneous magnetization $M$ in terms of the spontaneous magnetic field
$\boldsymbol{H}(\boldsymbol{r})$ as 
\begin{align}
	\boldsymbol{M} = 
  \frac{1}{\nu} 
	\int d \boldsymbol{r} 
	\boldsymbol{H}(\boldsymbol{r}). 
	\label{eq:Magnetization}
\end{align}
where $\nu = \pi R^2 D$ is the volume of a small superconducting disk. 
{In this paper, we did not calculate the magnetic field in the three-dimension. We obtain
$\boldsymbol{H}$ by solving the Maxwell equation in the $x$-$y$ plane with the boundary condition
$H(x,y) = 0$ outside of the disk, and assume that the magnetic field is homogeneous in the $z$
direction [i.e., $\boldsymbol{H} (\boldsymbol{r}) = H(x,y) \hat{\boldsymbol{z}}$ with
$\hat{\boldsymbol{z}}$ being the unit vector].  }
We iterate the Eilenberger equation for the Green function and
the Maxwell equation for the vector potential to obtain the self-consistent solutions of
$\Delta_1(\boldsymbol{r})$, $\Delta_2(\boldsymbol{r})$, $\boldsymbol{A}(\boldsymbol{r})$, and
$\check{\Sigma}(\boldsymbol{r},i\omega_n)$.

We start all of the simulations with the initial condition $\Delta_1(\boldsymbol{r}) =
\Delta_2(\boldsymbol{r})=|\bar{\Delta}(T)|$ and $\boldsymbol{A}(\boldsymbol{r}) = 0$, where
$|\bar{\Delta}(T)|$ is the amplitude of the pair potential in a homogeneous superconductor at a
temperature $T$.  Throughout this paper, we fix several parameters: the radius of a disk
$R=10\xi_0$, the cutoff energy $\omega_c=6 \pi T_c$. The magnetic field and the spontaneous
magnetization are measured in units of the second critical magnetic field $H_{c_2} = \hbar c /
|e| \xi_0^2$. 
{The current density is normalized to $j_0 = 2 |e| v_F N_0 T_c = \hbar c^2 / 4
\pi^2 |e| \lambda_L^2 \xi_0$.}
In the quasiclassical theory, the London length $\lambda_L = (mc^2/4\pi n_e e^2)^{1/2}$ with
$n_e$ being the electron density is a parameter characterizing the spatial variation of
magnetic fields, and is fixed at $\lambda_L = 5\xi_0$. In this paragraph, we explicitly denoted
$\hbar$ and $c$ to avoid misunderstandings. 

To solve the Eilenberger equation in a disk geometry, we apply the Riccati parametrization to
the Green function \cite{N.Schopohl_PRB_1995,N.Schopohl_arXiv_1998,M.Eschrig_PRB_2009} and the
technique discussed in Ref.~\onlinecite{Y.Nagai_PRB_2012}.  By using the Riccati
parametrization, we can separate the Eilenberger equation into the two Riccati-type
differential equations.  Solving the Riccati equations along a long enough quasiclassical
trajectory (typically 30 times of the coherence length), we can obtain the solutions of the
Eilenberger equation.  

\begin{figure*}[t]
	\begin{center}
	\includegraphics[width=0.75\textwidth]{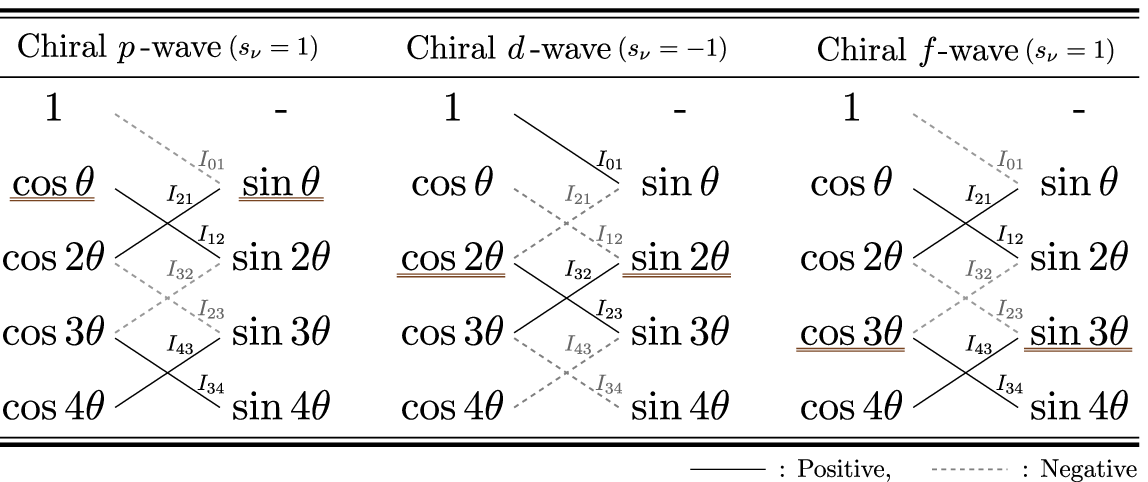}
  \caption{
    Chart of $I_{ab}$ in Eq.~(\ref{eq:Iab}). 
    The diagonal lines connecting $\cos (a\theta)$ and $\sin (b\theta)$ indicate possible combinations of 
    $f^{\mathrm{c}}_{a}$ and $ f^{\mathrm{s}}_{b} $ for the chiral currents. 
    The solid (broken) lines means $I_{ab}$ is positive (negative). 
	The sign of $I_{ab}$ in a spin-singlet superconductor is opposite to that in a spin-triplet superconductor
	due to an extra sign factor $s_\nu$. 
    The double underlines indicate the principal pairing component linking to the pair potential
    (i.e., $\cos ( \chi \theta )$ and $\sin (\chi \theta)$). At the first line, ``1'' represents 
	$s$\,-wave component.
  }
	\label{fig:chart}
  \end{center}
\end{figure*}

As we will demonstrate in the following sections, the edge currents show complicated spatial
profiles depending on the pairing symmetry.  To analyze such behaviors, we decompose the
electric current into a series of current components in terms of the symmetry of Cooper pairs.
By using the normalization relation $g^2 -  s_\nu f \undertilde{f}=1$ under the assumption $ f
\hspace{-0.5mm} \undertilde{f} \ll 1 $, we represent the normal Green function as $ {g} \approx
1 + s_\nu {f}\hspace{-0.5mm}\undertilde{f} / 2 $, where we have used the Eq.~(\ref{eq:spin_gf})
and $s_\nu=1$ ($s_\nu=-1$) for the spin-triplet (spin-singlet) pair potential.  By substituting
the expression into the current formula in Eq.~(\ref{eq:j}), the electric current can be
expressed as \cite{S.Higashitani_JPSJ_2014}
\begin{align}
 &\boldsymbol{j}(\boldsymbol{r})=
 \hspace{-1mm}\sum_{\omega_n>0} \boldsymbol{j}_{\omega_n}(\boldsymbol{r}) \\
 &\boldsymbol{j}_{\omega_n} =
 {4 \pi e v_F N_0} T \hspace{-1mm}
 \int \hspace{-1mm}\frac{ d \theta }{2 \pi } \,
 \frac{1}{2} s_\nu \bm{k} \,
 \mathrm{Im}[
 f  
 \hspace{-0.5mm}
 \undertilde{ f}],
 \label{eq:j2}
\end{align}
where we have used the relation $g(\boldsymbol{r}, \theta, i\omega_n) = -g^*(\boldsymbol{r},
\theta, -i\omega_n) $.  Generally speaking, the pairing function $f(\boldsymbol{r}, \theta,
i\omega_n)$ can be decomposed into the Fourier series
\begin{align}
	f(\boldsymbol{r}, \theta, i\omega_n) =& 
	\sum_{a=0} 
	f^{\mathrm{c}}_{a} 
	(\boldsymbol{r}, i\omega_n) \cos ( a \theta ) \notag \\ 
	& \hspace{5mm} 
	+\sum_{b=1} 
	i f^{\mathrm{s}}_{b} 
	(\boldsymbol{r}, i\omega_n) \sin ( b \theta ). 
\end{align}
The surface breaks locally the inversion symmetry and induces subdominant pairing components
whose symmetries are different from that of the pair potential.  In the absence of the vector
potential, $f^{\mathrm{c}}_{a}$ and $f^{\mathrm{s}}_{b}$ are real functions.  When we consider
the current profile at $y=0$, the electric current in the $y$ direction becomes
\begin{align}
  &{j}_{y}(x)= \, \sum_{\omega_n}
  \sum_{ab} j_{ab}(i\omega_n) \label{eq:jy}\\
  &{j}_{ab}(i\omega_n) = \,
  {4 \pi |e| v_F N_0}\, T\, 
  f^{\mathrm{c}}_{a} \, 
  f^{\mathrm{s}}_{b} \,
  I_{ab}
  \label{eq:jab} \\[2.5mm]
  &{I}_{ab} = \,
	s_\nu (-1)^b
	( \delta_{b,1-a}
	+ \delta_{b,a+1}
	- \delta_{b,a-1} ) / 4
  \label{eq:Iab} 
\end{align}
where we use the relation 
$ \undertilde{f}( \boldsymbol{r}, \theta, i\omega_n) = 
f^*( \boldsymbol{r}, \theta+\pi, i\omega_n)$, 
and $\int d\theta \sin \theta \cos (a\theta) \sin (b\theta) = 
( \delta_{b,1-a} + \delta_{b,a+1} - \delta_{b,a-1} )\pi/2$ 
for $a \ge 0$ and $b \ge 1$.  The Kronecker's $\delta$ functions appearing in
Eq.~(\ref{eq:Iab}) suggest that only the limited combinations of $f^{\mathrm{c}}_{a}$ and
$f^{\mathrm{s}}_{b} $ contribute to the supercurrents, (e.g., $a= b \pm 1$). 
Moreover, the direction of the decomposed current $j_{ab}$ in Eq.~(\ref{eq:jab}) depends on the
signs of $ f^{\mathrm{c}}_{a} f^{\mathrm{s}}_{b} $ and $I_{ab}$.  The
signs of $I_{ab}$ mainly determine the current directions because $ f^{\mathrm{c}}_{a}
f^{\mathrm{s}}_{b} $ appearing at a certain surface have the same signs in most cases.  We show a chart of
$\mathrm{sgn}[I_{ab}]$ in Fig.~\ref{fig:chart}.  The diagonal lines connecting $\cos (a\theta)$
and $\sin (b\theta)$ mean the possible combinations for carrying the currents.  The solid
(broken) lines indicate that $I_{ab}$ is positive (negative).  In a chiral $p$\,-wave
superconductor, for example, $I_{01}$ and $I_{21}$ have the opposite signs to each other. As a
result, the decomposed currents $j_{01}$ and $j_{21}$ flow in opposite directions.

\section{NON-SELF-CONSISTENT SIMULATION}

\begin{figure}[t]
	\begin{center}
	\includegraphics[width=0.45\textwidth]{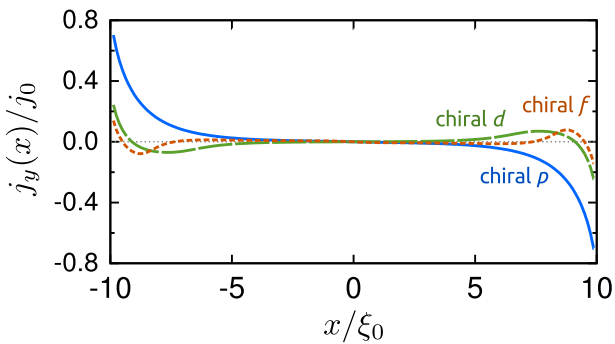}
	\caption{Current densities in a disk of a chiral superconductor with 
	a constant pair potential $\Delta_{1(2)}(\boldsymbol{r}) = |\bar{\Delta}(T)|$ at 
	 $\boldsymbol{A}
		(\boldsymbol{r})=0$, 
		where $|\bar{\Delta}(T)|$ is the amplitude of the pair potential at a temperature $T$ 
		in a homogeneous superconductor. 
	   Here, we show the current distribution at $y=0$. 
	  The radius of a superconducting disk, the temperature, and the cutoff energy 
    are set to $R=10\xi_0$, $T=0.2T_c$, and $\omega_c = 6 \pi T_c$. }
	\label{fig:cur_nsc}
  \end{center}
\end{figure}

Before turning into the effects of surface roughness and those of the Meissner screening, the
chiral currents in the uniform pair potential at $\boldsymbol{A}=0$ should be summarized.  The
results presented in this section are qualitatively the same as those obtained by the
Bogoliubov-de Gennes (BdG) formalism in Refs.~\onlinecite{Y.Tada_PRL_2015} and
\onlinecite{W.Huang_PRB_2014}. 

The spatial dependences of the edge current are shown in Fig.~\ref{fig:cur_nsc}, where we show
the spatial distribution of the current in the $y$ direction $j_y(x)$ at $y=0$, where the
temperature is set to $T = 0.2T_c$. The results are circular symmetric on a superconducting
disk. In a chiral $p$\,-wave superconductor ($\chi$ = 1), the amplitude of the edge current
takes its maximum at $r=R$ and monotonically decreases with increasing the distance from the
surface.  When we observe the current from the $+z$ axis, the chiral current flows in the
clockwise direction. 
The current distributions in chiral $d$\,-wave ($\chi=2$) and chiral $f$-wave ($\chi=3$)
superconductors are rather complicated than that in a chiral $p$\,-wave case.  The current
density is negative (clockwise) around $x/\xi_0=10$ and is positive (counterclockwise) for
$x/\xi_0<9$ in a chiral $d$\,-wave superconductor.  In a chiral $f$-wave case, the current
density is negative for $ 9.5 < x/\xi_0 < 10$, positive for $7.8 < x/\xi_0< 9.5$, and negative
again for $x/\xi_0<7.8$. 
The net current density $J=\int_0^R dx \, j_y(x)|_{y=0}$ decreases with 
increasing the chiral index $\chi$ because there are two (three) current channels in a
chiral $d$\,-wave ($f$-wave) superconductor and they carry the currents in opposite directions.  

\section{Disk with a specular surface}

In this section, we discuss the current distribution of a chiral-superconducting disk
with a specular surface under the self-consistent pair potentials and the vector potential. 
The results are obtained by solving the Eilenberger and Maxwell equations 
simultaneously and self-consistently. 
In Sec.~IV A, we consider only the self-consistent pair potential 
at $\boldsymbol{A}=0$ in Eq.~(\ref{H_def}) to analyze 
the complicated spatial distribution of the chiral current.
The results tell us the symmetry of Cooper pairs that carry the chiral current. 
The effects of self-induced magnetic fields are briefly discussed in Sec.~IV B.
The parameters are set to the same values used in Fig.~\ref{fig:cur_nsc}.


%
\subsection{Results under self-consistent pair potential at $\boldsymbol{A}= 0$ }

\begin{figure}[tb]
	\begin{center}
	\includegraphics[width=0.35\textwidth]{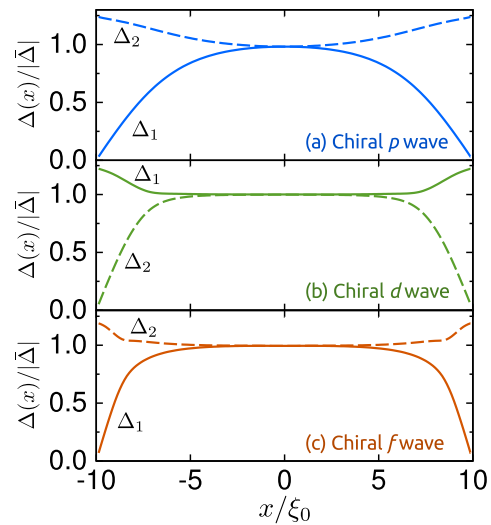}
	\caption{Pair potentials in a disk of a chiral superconductor. The results are obtained 
    by solving self-consistently the Eilenberger equation under the condition 
		$\boldsymbol{A} (\boldsymbol{r} ) = 0$. 
		The superconducting disk is in the clean limit. 
		The pair potentials are normalized to $|\bar{\Delta}(T)|$, the amplitude of the pair
    potential in a homogeneous superconductor at a temperature $T$. 
		The parameters are set to the same values used in Fig.~\ref{fig:cur_nsc}. }
	\label{fig:del_cln_A0}
  \end{center}
\end{figure}

\begin{figure}[tb]
	\begin{center}
	\includegraphics[width=0.40\textwidth]{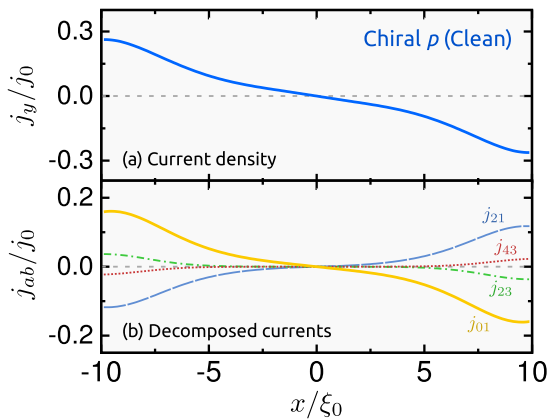}
	\caption{
    Results for a chiral $p$\,-wave disk with a {\it specular} surface obtained by the
    self-consistent simulation under $\boldsymbol{A}(\boldsymbol{r})=0$.  The chiral current
    $j_{y}(x)$ in Eq.~(\ref{eq:jy}) at $y=0$ is shown in (a).  The decomposed current 
    $j_{ab}$ at the lowest Matsubara frequency in Eq.~(\ref{eq:jab}) are shown in (b). 
    All of the currents in (a) and (b) are normalized to {$j_0 = 2 |e| v_F
    N_0 T_c$}. 
    The parameters are set to the same values used in Fig.~\ref{fig:cur_nsc}. 
  }
	\label{fig:curpair_cln_p}
  \end{center}
\end{figure}

%
In Fig.~\ref{fig:del_cln_A0}, we show the spatial dependence of the pair potentials $\Delta_1$
and $\Delta_2$.  In a chiral $p$\,-wave superconductor, the pair potential $\Delta_1$ is
strongly suppressed, whereas $\Delta_2$ is slightly enhanced near the surface as shown in
Fig.~\ref{fig:del_cln_A0}(a).  These suppression and enhancement are closely related to the
formation of the surface ABSs. Namely, $\Delta_1$ changes its sign while the quasiparticle is
reflected by a specular surface.  These spatial variations of the pair potentials affect the
edge current. 
The current density $j_y(x)$ at $y=0$ in Eq.~(\ref{eq:jy}) is shown in
Fig.~\ref{fig:curpair_cln_p}(a).  In a chiral $p$\,-wave disk, the edge current monotonically
decreases with increasing the distance from the edge.  The edge current under the self-consistent
pair potential in Fig.~\ref{fig:curpair_cln_p}(a) flows much wider area than that obtained by
the uniform pair potential in Fig.~\ref{fig:cur_nsc}.  The range of ``edge'' is determined by
the spatial variation of the pair potential in Fig.~\ref{fig:del_cln_A0}.

The surface breaks locally the inversion symmetry and the spatial variation of the pair
potential breaks the translational symmetry. As a result, the subdominant pairing correlations
are induced near the surface. 
\cite{Comment} 
In Fig.~\ref{fig:chart}, we enumerate the orbital symmetry of
such subdominant components.  The double underlines indicate the principal pairing component
linked to the pair potential.  At the first row, ``1'' represents $s$\,-wave symmetry.  In a
chiral $p$\,-wave case, the spatial variation of the principal component $\cos(\theta)$ induces
the subdominant component such as $s$\,-wave, $d$\,-wave $\cos(2\theta)$, $f$-wave
$\cos(3\theta)$, $\cdots$.  In the same way, the principal component $\sin(\theta)$ induces the
subdominant component of $d$\,-wave $\sin(2\theta)$, $f$-wave $\sin(3\theta)$, $\cdots$.  The
current is decomposed into the series of $j_{ab}$ in Eq.~(\ref{eq:jab}).  
The results for a
chiral $p$\,-wave disk are shown in Fig.~\ref{fig:curpair_cln_p}(b), where $j_{01}$, $j_{21}$,
$j_{23}$ and $j_{43}$ contribute mainly to the current.  Here $j_{ab}$ shown in 
Fig.~\ref{fig:curpair_cln_p}(b) are calculated at the
lowest Matsubara frequency $\omega_0$. 
{We have confirmed that $\sum_{ab} j_{ab}(\omega_0)$ is
almost identical to the current density obtained from the normal Green function $j(\omega_0)$, 
and that the components at higher Matsubara frequencies have almost the similar spatial distribution as
$j_{ab}$ at $\omega_0$.  
}
Reflecting the signs of $I_{ab}$ in
Fig.~\ref{fig:chart}, $j_{01}$ and $j_{23}$ flow in the clockwise direction, whereas $j_{21}$
and $j_{43}$ do in the counterclockwise direction. The magnitudes of $j_{01}$ and $j_{23}$ are
slightly larger than $j_{21}$ and $j_{43}$, respectively. As a consequence, the net edge
current flows in the clockwise direction.  We have confirmed that another possible $j_{ab}$ are
negligible.  
The decomposed currents in Fig.~\ref{fig:curpair_cln_p}(b) 
tell us the symmetry of Cooper pairs that carry the edge current. 
The partial current $j_{01}$ represents the current carried by the combination 
of $s$\,-wave and $p_y$\,-wave Cooper pairs. The current $j_{21}$ are also understood as
the current carried by $d_{x^2-y^2}$-wave $\times$ $p_y$-wave Cooper pairs. 
 
All of the Cooper pairs in a chiral $p$\,-wave superconductor belong to the spin-triplet
symmetry class in the absence of spin-dependent potentials.  Therefore, even-parity pairs
induced at a surface have the odd-frequency symmetry because of the anti-symmetry relation
derived from the Fermi-Dirac statistics of electrons
\begin{align}
   \hat{f}              ( \boldsymbol{r}, \theta    , i\omega_n) = 
  -\hat{f}^{\mathrm{T}} ( \boldsymbol{r}, \theta+\pi,-i\omega_n), 
\label{eq:FD}
\end{align}
where $\cdot^{\mathrm{T}}$ represents the transpose of a matrix and means the commutation of
the two spins of a Cooper pair. The odd-parity symmetry accounts the negative sign on the
right-hand side of Eq.~(\ref{eq:FD}) in a spin-triplet superconductor.  On the other hand, the
induced spin-triplet even-parity components satisfy Eq.~(\ref{eq:FD}) by their frequency
dependence. They are so-called odd-frequency Cooper pairs. 
\cite{V.L.Berezinskii_JETP_1974,Y.Tanaka_JPSJ_2012} 
As shown in Fig.~\ref{fig:curpair_cln_p}(b) and Eq.~(\ref{eq:jab}), the spontaneous edge current
in a chiral superconductor is carried by the combination of the even- and odd-frequency Cooper
pairs staying at a surface.

\begin{figure}[tb]
	\begin{center}
	\includegraphics[width=0.40\textwidth]{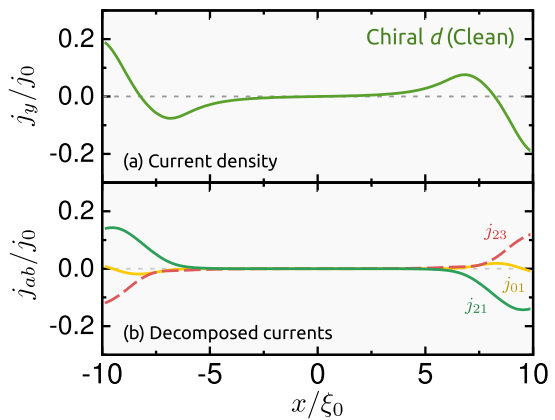}
	\caption{
    Results for a chiral $d$\,-wave disk with a specular surface obtained by the self-consistent simulation 
		under $\boldsymbol{A}(\boldsymbol{r})=0$:
	current density $j_{y}$ (a) and 
	dominant components $j_{ab}$ (b).
	The results are plotted in the same manner as Fig.~\ref{fig:curpair_cln_p}. 
  }
	\label{fig:curpair_cln_d}
  \end{center}
\end{figure}


In a chiral $d$\,-wave superconductor, $\Delta_2$ is responsible for the formation of the
surface ABSs. Correspondingly, the pair potential $\Delta_1$ is slightly enhanced near the
surface as shown in Fig.~\ref{fig:del_cln_A0}(b).  The spatial profile of the current  is shown
in Fig.~\ref{fig:curpair_cln_d}(a).  As is the case in the non-self-consistent simulation,
there are two edge channels in a chiral $d$\,-wave disk.  The current in the clockwise
direction flows along the surface and the current in the counterclockwise flows around $x=\pm
7\xi_0$.  In Fig.~\ref{fig:curpair_cln_d}(b), we decompose the current into the series of
$j_{ab}$, where we show only dominant components of $j_{21}$, $j_{23}$ and $j_{01}$.  We note
that $j_{12}$ and $j_{32}$ (not shown) have almost the same profile as $j_{01}$, and that another
components are negligible.  The principal pairing components in a chiral $d$\,-wave
superconductor are $f^{\mathrm{c}}_{2}\cos(2\theta)$ and $f^{\mathrm{s}}_{2}\sin(2\theta)$ as
shown in Fig.~\ref{fig:chart}.  The spatial variation of the pair potential generates the
odd-frequency components $f^{\mathrm{s}}_{1}\sin(\theta)$ and
$f^{\mathrm{s}}_{3}\sin(3\theta)$.  These induced components carry the spontaneous current
indicated by $j_{21}$, $j_{23}$, $j_{12}$ and $j_{32}$.  As shown in
Fig.~\ref{fig:curpair_cln_d}(b), $j_{21}$ and $j_{23}$ flow in opposite directions 
because $I_{21}$ and $I_{23}$ have opposite signs. As a result, the
net edge current becomes smaller than that in a chiral $p$\,-wave disk. 

\begin{figure}[tb]
	\begin{center}
	\includegraphics[width=0.40\textwidth]{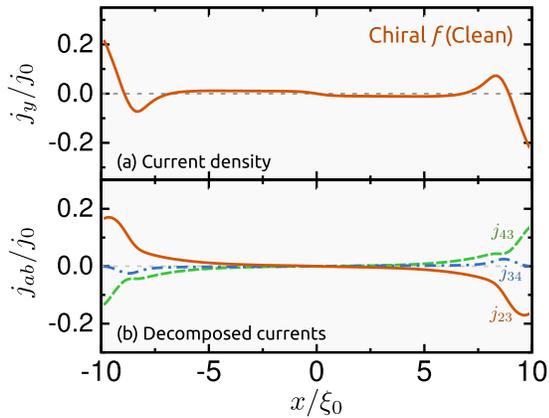}
	\caption{
    Results for a chiral $f$\,-wave disk with a specular surface obtained by the self-consistent simulation 
		under $\boldsymbol{A}(\boldsymbol{r})=0$:
	current density $j_{y}$ (a) and 
	dominant components $j_{ab}$ (b).
	The results are plotted in the same manner as Fig.~\ref{fig:curpair_cln_p}. 
  }
	\label{fig:curpair_cln_f}
  \end{center}
\end{figure}

In a chiral $f$-wave disk, $\Delta_1$ is suppressed and $\Delta_2$ is slightly enhanced near
the surface due to the emergence of the surface ABSs as shown in Fig.~\ref{fig:del_cln_A0}(c).
The current profile and the decomposed currents $j_{ab}$ are shown in
Fig.~\ref{fig:curpair_cln_f}(a) and \ref{fig:curpair_cln_f}(b), respectively.  Although the
spatial profile of the current is greatly modified by the self-consistent pair potentials,
Fig.~\ref{fig:curpair_cln_f}(a) suggests that there are three current channels.
The current density is negative for $ 8 < x/\xi_0 $, is positive for $ 6<x/\xi_0 < 8$, and 
is negative again for $ 0 < x/\xi_0 < 6$. 
Figure.~\ref{fig:curpair_cln_f}(b) shows that  the spatial dependence of the current components
$j_{23}$, $j_{43}$ and $j_{34}$ are responsible for such a complicated current profile.  We note
that $j_{21}$ and $j_{32}$ (not shown) have almost the same profile as $j_{34}$.

\subsection{Results under self-consistent pair potential and vector potential}

\begin{figure}[tb]
	\begin{center}
	\includegraphics[width=0.40\textwidth]{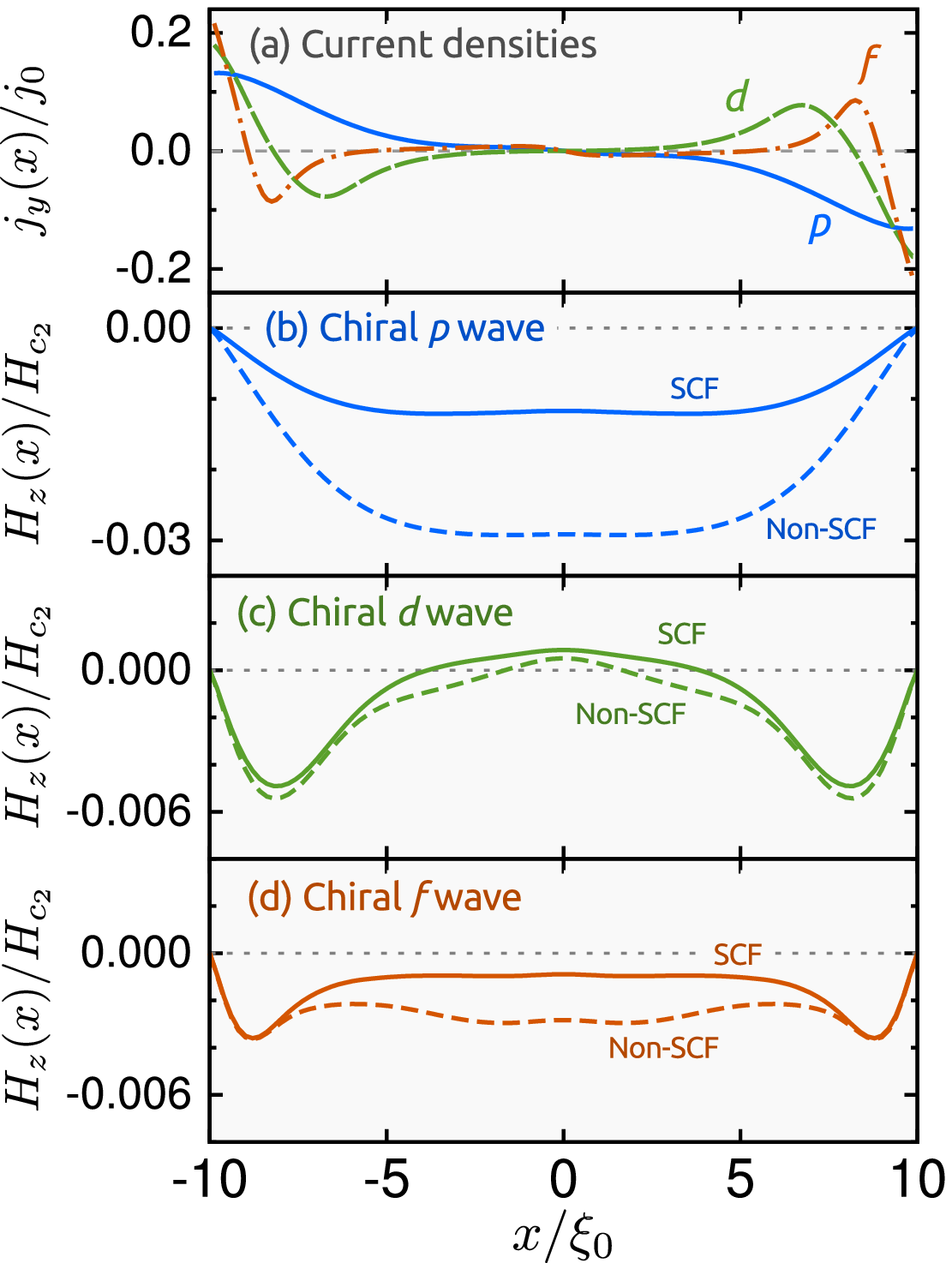}
  \caption{(a) Current densities in a disk of a chiral superconductor with a {\it specular} surface (a).
    The results are obtained by solving the Eilenberger and Maxwell equations self-consistently
    and simultaneously.  The penetration depth is fixed at $\lambda_L = 5 \xi_0$.  The other
    parameters are set to the same values used in Fig.~\ref{fig:cur_nsc}.  The current
    densities are normalized to {$j_0 = 2 |e| v_F N_0 T_c$}. 
    In (b)-(d), we compare the spatial
    distributions of the self-consistent fields (SCF) with those of the non-self-consistent
    fields (non-SCF). The former is obtained by the current densities in (a) by using the
    relation in Eq.~(\ref{mag_j}). The latter is calculated from the current distributions in
    Figs.~\ref{fig:curpair_cln_p}(a), \ref{fig:curpair_cln_d}(a), and
    \ref{fig:curpair_cln_f}(a). 
    The magnetic fields are scaled in units of $H_{c_2} = \hbar c / |e| \xi_0^2$. 
 }
	\label{fig:curmag_cln}
  \end{center}
\end{figure}

\begin{figure}[tb]
	\begin{center}
	\includegraphics[width=0.35\textwidth]{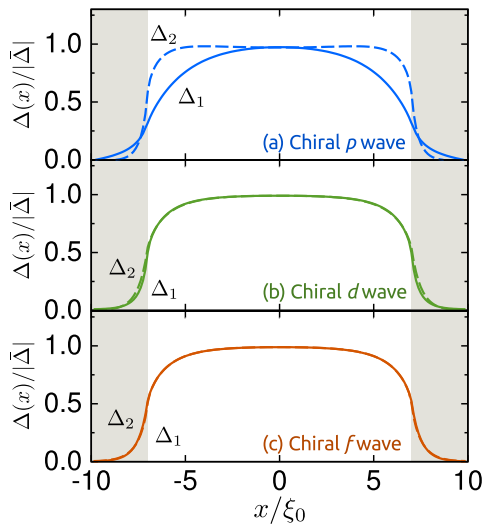}
	\caption{Pair potentials in a disk of a chiral superconductor with a rough surface as indicated by 
	the shadowed area. The results are obtained 
		by solving the Eilenberger equation self-consistently at $\boldsymbol{A}=0$. 
		The parameters are set to the same values used in Fig.~\ref{fig:cur_nsc}. 
	The magnetic penetration depth is $\lambda_L = 5 \xi_0$. 
	}
	\label{fig:del_rgh_A0}
  \end{center}
\end{figure}

We take into account the vector potential $\boldsymbol{A}$ induced by the edge current to
investigate the Meissner screening effect.  The pair potential and the vector potential are
determined in a self-consistent way by solving the Eilenberger and Maxwell
equations simultaneously.  The spatial profiles of the pair potentials are qualitatively the same
as those in Fig.~\ref{fig:del_cln_A0}.  The spatial profiles of the chiral edge currents are
shown in Fig.~\ref{fig:curmag_cln}(a).  In Figs.~\ref{fig:curmag_cln}(b)-\ref{fig:curmag_cln}(d), we compare the
local magnetic fields obtained under the self-consistent field (SCF) with that under the
non-self-consistent field of the vector potential (non-SCF).  The latter is calculated from the
current distribution in Figs.~\ref{fig:curpair_cln_p}(a), \ref{fig:curpair_cln_d}(a), and
\ref{fig:curpair_cln_f}(a) by using the relation in Eq.~(\ref{mag_j}). 

In a chiral $p$\,-wave disk, the Meissner screening by the superconducting condensate
suppresses dramatically the spontaneous magnetization as shown in Fig.~\ref{fig:curmag_cln}(b).  As a
result, the amplitude of the current at $x=R$ is less than {$0.13 j_0$} under the SCF in
Fig.~\ref{fig:curmag_cln}(a), whereas it is about {$0.27j_0$} under the non-SCF in
Fig.~\ref{fig:curpair_cln_p}(a). The magnetic field near the center of a disk remains at a 
finite value in both the SCF and non-SCF simulations. 
This magnetic-field penetration is
a results of the finite-size effects. At
the surface of a semi-infinite sample, we have confirmed the
current inversion because of the Meissner screening current as seen in
Fig.~2 in Ref.~\onlinecite{M.Matsumoto_JPSJ_1999}. Namely, the bulk condensate generates the
screening current which flows in the opposite direction to the chiral current at the
surface. As a result, the magnetic field in the bulk region vanishes in a semi-infinite
superconductor.

In a chiral $d$\,-wave superconductor, the magnetic field is mainly localized around $x=\pm 8
\xi_0$ as shown in Fig.~\ref{fig:curmag_cln}(c). The results with the SCF is slightly smaller
than those with the non-SCF.  Thus the Meissner effect in a chiral $d$\,-wave disk is much
weaker than that in a chiral $p$\,-wave one.  The current profile under the non-SCF in
Fig.~\ref{fig:curpair_cln_d}(a) shows that there are two channels for the edge current. One is the outer channel
for  the clockwise current and the other is the inner channel for the counterclockwise current.
The induced magnetic field by the inner current well screens that by the outer current
intrinsically. Such a self-screening effect makes the Meissner screening effect weak in a chiral
$d$\,-wave disk. Actually such characteristic current profile with the non-SCF in
Fig.~\ref{fig:curpair_cln_d}(a) are well preserved in the results with the SCF in
Fig.~\ref{fig:curmag_cln}(a).  The current amplitude at the surface $x=R$ reaches to about
{$0.18j_0$}
in Fig.~\ref{fig:curmag_cln}(a) and it is about 
{$0.19j_0$}
in Fig.~\ref{fig:curpair_cln_d}(a). Thus, in a chiral $d$\,-wave disk, the Meissner effect
modifies the edge current only slightly as shown in Fig.~\ref{fig:curmag_cln}(c). 

The result of the edge current for a chiral $f$-wave superconductor in
Fig.~\ref{fig:curmag_cln}(d) can be explained in the same way. There are three channels for the
edge current in a chiral $f$-wave case as discussed in Fig.~\ref{fig:curpair_cln_f}(a). The
self-screening effect works in this case as well. The characteristic behavior of the edge
current with the non-SCF in Fig.~\ref{fig:curpair_cln_f}(a) remain almost unchanged even
with the SCF as shown in Fig.~\ref{fig:curmag_cln}(a). However, because the self-screening
effect does not sufficiently exclude the local field, the magnetic field around the
center of a disk is suppressed by the Meissner effect as shown in Fig.~\ref{fig:curmag_cln}(d).

\begin{figure}[tb]
	\begin{center}
	\includegraphics[width=0.40\textwidth]{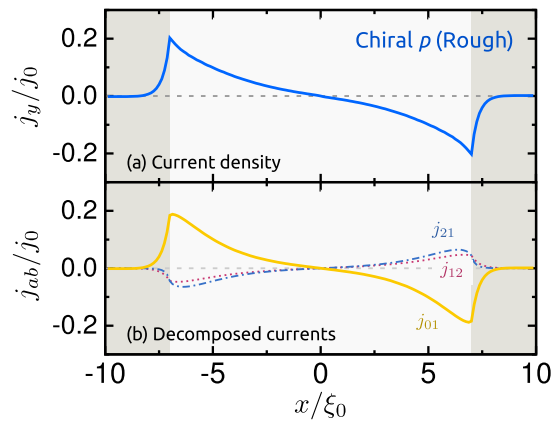}
	\caption{
    Results for a chiral $p$\,-wave disk with a rough surface obtained by the self-consistent simulation 
		at $\boldsymbol{A}(\boldsymbol{r})=0$. 
	  The current density $j_{y}(x)$ at $y=0$ in Eq.~(\ref{eq:jy}) is shown in (a). 
    The dominant components $j_{ab}$ at the lowest Matsubara frequency are shown in (b).
  }
	\label{fig:curpair_rgh_p}
  \end{center}
\end{figure}

\section{Disk with a rough surface}

In this section, we discuss the effects of a rough surface on the chiral edge currents. 
The width of the disordered region [shadowed in Fig.~\ref{fig:sche}(b)]
is set to be $W=3\xi_0$ because the chiral 
edge current in the clean limit mainly flows in such area as shown in Fig.~\ref{fig:curmag_cln}(a). 
The strength of roughness is set to $\xi_0 / \ell = 1.0$, where $\ell=v_F \tau_0 $ is the elastic 
mean free path of a quasiparticle. 
The another parameters are set to the same values used in Fig.~\ref{fig:curpair_cln_p}. 
The rough surface drastically changes the spatial profile of the pair potential
and that of induced subdominant pairing components.
Thus we first summarize symmetry of Cooper pairs appearing near the rough surface 
in Sec.~V A. Then we discuss briefly the Meissner screening effect by the bulk condensate in Sec.~V B.
%
\subsection{Results under self-consistent pair potential at $\boldsymbol{A}=0$}

Here we discuss the results obtained by solving {\it only} the Eilenberger equation 
under the condition 
$\boldsymbol{A}=0$ in Eq.~(\ref{H_def}). 
We obtain the self-consistent solutions of $\Delta_1(\boldsymbol{r})$, 
$\Delta_2 (\boldsymbol{r})$, and $\hat{\Sigma} (\boldsymbol{r}, i\omega_n)$. 
The pair potentials are presented in Fig.~\ref{fig:del_rgh_A0}.
The surface roughness 
strongly suppresses the pair potentials $\Delta_1$ and $\Delta_2$ in the disordered region of a
chiral $p$\,-wave disk. 
At the interface between the disordered and clean regions (we refer to it as the d/c interface
in what follows), $\Delta_1$ is suppressed more significantly than $\Delta_2$, which
suggests the formation of the ABSs there.~\cite{S.V.Bakurskiy_PRB_2014}
We show the current density $j_{y}$ at $y=0$ and 
the dominant current components $j_{ab}$ in Fig.~\ref{fig:curpair_rgh_p}.
Comparing Fig.~\ref{fig:curpair_cln_p}(a) with \ref{fig:curpair_rgh_p}(a), 
one can find that the peak of the edge current moves from the surface to the d/c interface, 
and that its maximum value 
{$0.20j_0$}
is comparable to the maximum value in the clean limit. 
As shown in Fig.~\ref{fig:curpair_rgh_p}(b), the edge current in a chiral $p$\,-wave disk is
mainly carried by three components; $j_{01}$, $j_{21}$, and $j_{12}$. Among them, the
combination of $s$\,-wave $\times$ $p_y$-wave pairs ($j_{01}$) dominates obviously the chiral current in a
disk with a rough surface. 
The spatial variation in $\Delta_1$ generates the $s$\,-wave and $d_{x^2-y^2}$-wave
odd-frequency pairs.\cite{S.I.Suzuki_PRB_2015} The induced $s$\,-wave pairs, in particular, are robust even
under the random potential. Such property supports the robustness of the chiral edge current
in a chiral $p$\,-wave superconductor. 

\begin{figure}[tb]
	\begin{center}
	\includegraphics[width=0.40\textwidth]{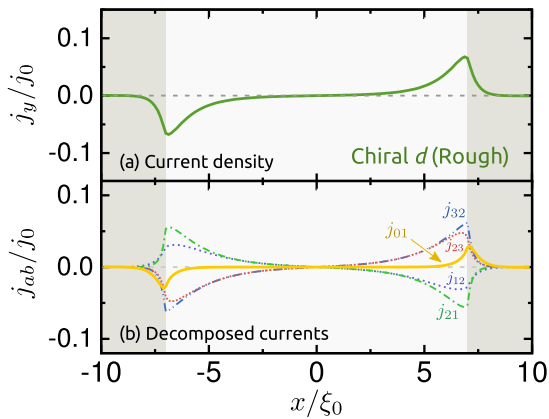}
	\caption{
    Results for a chiral $d$\,-wave disk with a rough surface obtained by the self-consistent simulation 
		at $\boldsymbol{A}(\boldsymbol{r})=0$:
	current density $j_{y}$ (a) and 
	dominant components $j_{ab}$ (b).
	The results are plotted in the same manner as Fig.~\ref{fig:curpair_rgh_p}. 
  }
	\label{fig:curpair_rgh_d}
  \end{center}
\end{figure}

The edge current in a chiral $d$\,-wave disk shows a qualitatively different behavior from that
in a chiral $p$\,-wave case.  As shown in Fig.~\ref{fig:curpair_rgh_d}(a), the chiral current
in a disk with surface roughness flows only in the counterclockwise direction.  This behavior
can be understood by comparing the current profile in Fig.~\ref{fig:curpair_cln_d}(a) with
that in Fig.~\ref{fig:curpair_rgh_d}(a).  In the clean disk, there are two edge currents: the
outer current running flowing in the clockwise direction and the inner current running in the
counterclockwise direction as shown in Fig.~\ref{fig:curpair_cln_d}(a).  The surface roughness
eliminates the outer current channel.  However, the inner current channel remains even in the
presence of the surface roughness and are responsible for the chiral current in the
counterclockwise direction.  We have confirmed that the inner current can survive in the
presence of much stronger roughness such as $\xi_0 / \ell \sim 30$. 
The decomposed components of the current $j_{ab}$ are shown in Fig.~\ref{fig:curpair_rgh_d}(b).
The edge current is mainly carried by five combinations: $j_{01}$, $j_{12}$, $j_{21}$,
$j_{23}$, and $j_{32}$.  The four components $j_{12}$, $j_{21}$, $j_{23}$, and $j_{32}$ almost
cancel one another. 
As shown in Fig.~\ref{fig:del_rgh_A0}(b),
the surface roughness suppresses both $\Delta_1$ and $\Delta_2$ in the same manner near 
the shadowed area, which results in 
\begin{align}
f^\mathrm{c}_2(x) \simeq f^\mathrm{s}_2(x).\label{d_relation0}
\end{align}
The spatial variation of 
$\Delta_1$ generates mainly $f^\mathrm{c}_1\cos(\theta)$ and $f^\mathrm{c}_3\cos(2\theta)$ 
with $f^\mathrm{c}_1(x) \simeq f^\mathrm{c}_3(x)$. 
In the same way, the spatial variation of
$\Delta_2$ induces $f^\mathrm{s}_1\sin(\theta)$ and $f^\mathrm{s}_3\sin(3\theta)$ with 
$f^\mathrm{s}_1(x) \simeq f^\mathrm{s}_3(x)$.
Therefore, the relation 
\begin{align}
f^\mathrm{c}_1(x) \simeq f^\mathrm{c}_3(x) \simeq f^\mathrm{s}_1(x) \simeq f^\mathrm{s}_3(x),
\label{d_relation}
\end{align}
holds among the four coefficients.
By applying the relation in Esq.~(\ref{d_relation0})
and (\ref{d_relation}) into Eq.~(\ref{eq:jab}) with the $I_{ab}$ in Fig.~\ref{fig:chart}, 
we can conclude that $j_{21}$ cancels $j_{32}$, and $j_{12}$ cancels $j_{23}$.  
The remaining component $j_{01}$, the contribution from the $s$\,-wave $\times$ $p_y$-wave pairs,
dominates the edge current.  
Because $s$\,-wave Cooper pairs are robust against surface roughness, 
$j_{01}$ can exist even under much stronger disordered potential.

\begin{figure}[tb]
	\begin{center}
	\includegraphics[width=0.40\textwidth]{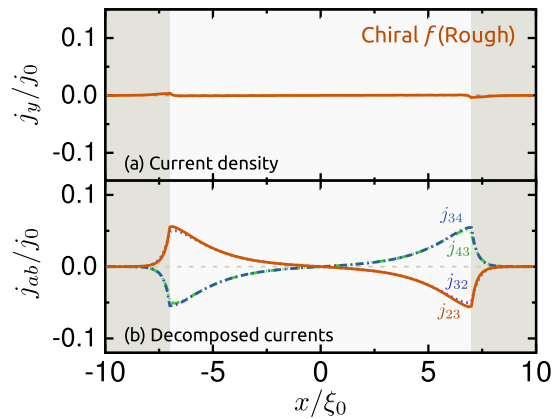}
	\caption{
    Results for a chiral $f$\,-wave disk with a rough surface obtained by the self-consistent simulation 
		at $\boldsymbol{A}(\boldsymbol{r})=0$:
	current density $j_{y}$ (a) and 
	dominant components $j_{ab}$ (b).
	The results are plotted in the same manner as Fig.~\ref{fig:curpair_rgh_p}.
	} 
	\label{fig:curpair_rgh_f}
  \end{center}
\end{figure}
As shown in Fig.~\ref{fig:curpair_rgh_f}(a), the edge current in a chiral
$f$-wave disk with a rough surface becomes almost zero in this scale of the plot (i.e.,
$|j_{y}| \ll j_0$). Within the accuracy of our numerical
simulation, the maximum value of the current density is less than {$4 \times
  10^{-3} j_0$}.
The dominant components $j_{23}$, $j_{32}$, $j_{34}$, and $j_{43}$ are shown in
Fig.~\ref{fig:curpair_rgh_f}(b).
As shown in Fig.~\ref{fig:del_rgh_A0}(c),
the surface roughness suppresses both $\Delta_1$ and $\Delta_2$ in the same manner near 
the shadowed area. By applying the same logic used in a chiral $d$\,-wave case, it is possible to 
show the relations
\begin{align}
&f^\mathrm{c}_3(x) \simeq f^\mathrm{s}_3(x),\\
&f^\mathrm{c}_2(x) \simeq f^\mathrm{c}_4(x) \simeq f^\mathrm{s}_2(x)  \simeq f^\mathrm{s}_4(x).
\label{f_relation}
\end{align}
These relations and $I_{ab}$ in Fig.~\ref{fig:chart} explain the cancellation among the current 
components such as $j_{23}+j_{34} \simeq 0$ and $j_{32}+j_{43} \simeq 0$.
As a result, the net edge current totally disappears as shown in Fig.~\ref{fig:curpair_rgh_f}(a).

The symmetry of Cooper pairs is determined by the pair potential and the random impurity
potential at a surface. Thus, even if a superconductor is semi-infinitely large and is realized
with a single chiral domain, we can find the similar behavior of the edge currents against
surface roughness as they show in a small superconductor. 

\begin{figure}[tb]
	\begin{center}
	\includegraphics[width=0.40\textwidth]{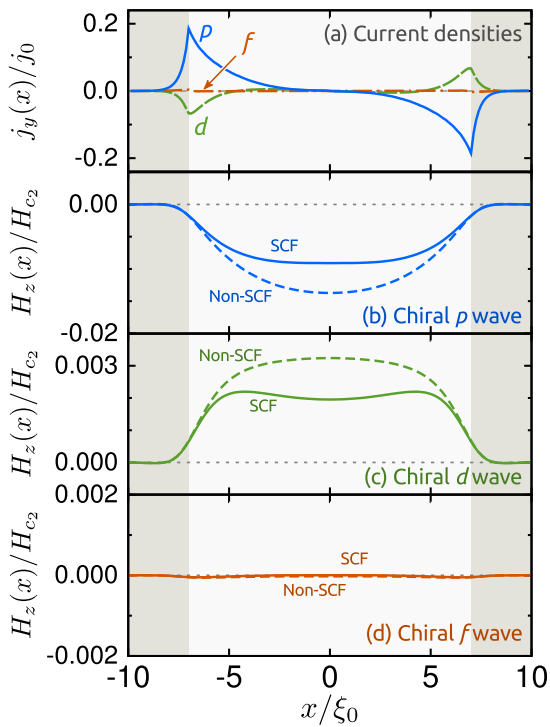}
  \caption{(a) Current densities in a disk of a chiral superconductor with a {\it rough} surface. 
	The results are obtained by solving the Eilenberger and Maxwell equations  
		self-consistently and simultaneously. (b)-(d) Comparisons of the SCF with the non-SCF. 
		 The strength of the disorder and the width of 
		the disordered region is set to $\xi_0 / \ell = 1$ and $ W=3\xi_0$, respectively. 
		The another parameters are set to the same values used in Fig.~\ref{fig:curmag_cln}. 
}
	\label{fig:curmag_rgh}
  \end{center}
\end{figure}


\subsection{Results under self-consistent pair potential and vector potential}

We discuss the effects of the self-induced vector potential on the chiral current in a disk
with surface roughness.  By solving simultaneously the Eilenberger and Maxwell equations, we
obtain the self-consistent solutions of $\Delta_1 ( \boldsymbol{r})$, $\Delta_2(
\boldsymbol{r})$, $\boldsymbol{A}(\boldsymbol{r})$, and $\hat{\Sigma}(\boldsymbol{r},
i\omega_n)$.  Here we do not show the pair potentials because they remain unchanged from those
in Fig.~\ref{fig:del_rgh_A0} even quantitatively.  The results of the edge currents are shown
in Fig.~\ref{fig:curmag_rgh}(a).  The spatial distributions of the magnetic field are presented
in Figs.~\ref{fig:curmag_rgh}(b)-(d).  For comparison, we show the non-SCF calculated from the
current profiles of Figs.~\ref{fig:curpair_rgh_p}(a), \ref{fig:curpair_rgh_d}(a) and
\ref{fig:curpair_rgh_f}(a) by applying the relation in Eq.~(\ref{mag_j}).

As shown in Fig.~\ref{fig:curmag_rgh}(b), the Meissner effect suppresses the magnetic field
around the center of a chiral $p$\,-wave disk.  When we compare the results for a chiral
$p$\,-wave disk with the SCF in Fig.~\ref{fig:curmag_rgh}(a) and those with the non-SCF in
Fig.~\ref{fig:curpair_rgh_p}(a), the current profile in Fig.~\ref{fig:curmag_rgh}(a) is
spatially compressed into a narrower region by the Meissner effect. 

The similar features are found also in the results of a chiral $d$\,-wave disk as shown in
Fig.~\ref{fig:curmag_rgh}(c).  In the presence of the surface roughness, the current profile
under the non-SCF has a monotonic spatial dependence between the center of the disk and the d/c
interface as presented in Fig.~\ref{fig:curpair_rgh_d}(a). 
Therefore, the self-screening effect observed in a clean disk does not work at all.  In a
chiral $d$\,-wave disk with a rough surface, the Meissner effect becomes stronger than that in
a disk with a specular surface.  The Meissner effect suppress the magnetization near the center
of the disk as shown in Fig.~\ref{fig:curmag_rgh}(c).

In a chiral $f$\,-wave disk, the surface roughness strongly suppresses the chiral current. Thus
the magnetic field is much smaller than $H_{c_2}$ everywhere in the disk as shown in
Fig.~\ref{fig:curmag_rgh}(d).  Within the numerical accuracy, the magnetic field is less than
$7\times 10^{-5} H_{c_2}$ in our simulation.


\section{Temperature dependence of spontaneous magnetizations}

\begin{figure}[t]
  \vspace{10mm}
	\begin{center}
	\includegraphics[width=0.40\textwidth]{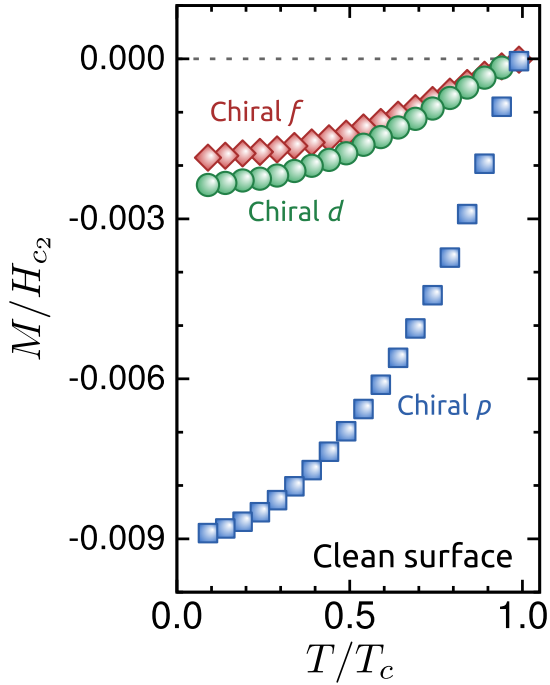}
	\caption{Temperature dependences of the spontaneous magnetization of a small chiral 
    superconductor with a {\it clean} surface.  
		The magnetization is defined by Eq.~(\ref{eq:Magnetization}) 
		and is normalized to the second critical magnetic field $H_{c_2}$. }
	\label{fig:chi_cln}
  \end{center}
\end{figure}

\begin{figure}[t]
  \vspace{10mm}
	\begin{center}
	\includegraphics[width=0.40\textwidth]{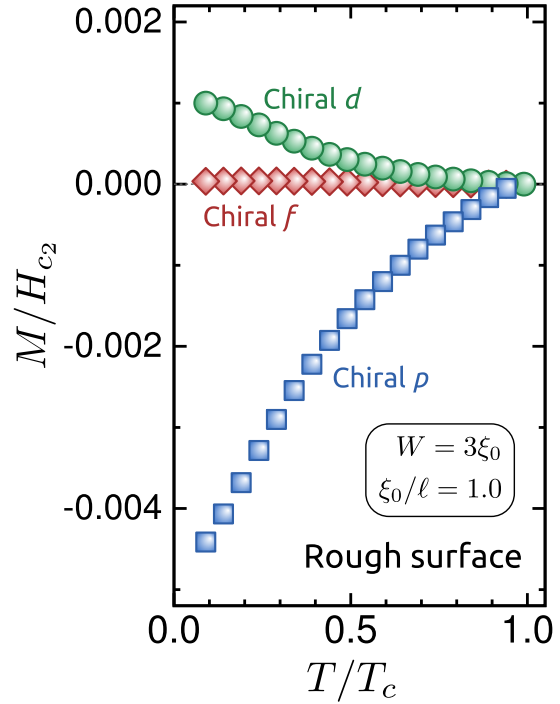}
	\caption{Temperature dependences of the spontaneous magnetization of a small chiral 
    superconductor with a {\it rough} surface. The parameters 
		related to the surface roughness are set to $\xi_0 / \ell = 1.0 $ and $W=3\xi_0$.  }
	\label{fig:chi_rgh}
  \end{center}
\end{figure}

Finally, we discuss the dependences of the spontaneous magnetization on temperature which are
measurable values in experiments. All of the simulations ware started at $T \simeq T_c$ with a
homogeneous pair potential $\Delta_1(\boldsymbol{r}) = \Delta_2(\boldsymbol{r}) = |
\bar{\Delta}(T) |$ and without an external magnetic field. The magnitude of a spontaneous
magnetization is defined in Eq.~(\ref{eq:Magnetization}).  In our simulations, the pair
potential, the impurity self-energy, and the vector potential are calculated self-consistently.
The results in a disk with a {\it specular} surface are shown in Fig.~\ref{fig:chi_cln}.  At a low
temperature $T= 0.1T_c$, the magnetization of a chiral $p$\,-wave disk reaches to about
$0.009H_{c_2}$. In chiral $d$\,- and $f$-wave disks, the magnetizations are about
$0.002H_{c_2}$. Although the magnetization decreases with increasing the radius of a disk by
its definition Eq.~(\ref{eq:Magnetization}), $0.002H_{c_2}$ at $R=10\xi_0$ would be detectable
value in experiments.  The results in a disk with a {\it rough} surface are shown in
Fig.~\ref{fig:chi_rgh}, we choose $\xi_0/\ell = 1.0 $ and $ W = 3 \xi_0 $.  In a chiral
$p$\,-wave superconductor, the amplitude of the magnetization is smaller than the results in
the clean limit at every temperatures.  As we discussed in Sec.~V, however, the amplitude of
the current density in a disk with a rough surface is comparable to that in a disk with a
specular surface.  In Eq.~(\ref{eq:Magnetization}), the magnetization is normalized by the area
of a whole disk. As shown in Fig.~\ref{fig:del_rgh_A0}, however, the effective radius of the
superconducting region shrinks down to $R_{\mathrm{eff}} = R-W$ in the presence of the
surface roughness.  When we renormalize the magnetization by the effective superconducting area, the
renormalized magnetization $\tilde{M} = M R^2/(R-W)^2 \approx 2 M$ is comparable to the
magnetization in the clean disk. This fact means the robustness of the chiral current in the
presence of the surface roughness.

In a chiral $d$\,-wave disk, the sign of the magnetization in Fig.~\ref{fig:chi_rgh} changes
from that in Fig.~\ref{fig:chi_cln} because only the inner chiral edge channel survives in a
disk with a rough surface and flows the current in the counterclockwise direction.  We have
confirmed that the magnetization of a chiral $d$\,-wave disk becomes small but remains finite
even in the presence of the much stronger disorder (e.g., $\xi_0/\ell = 30 $). As discussed in
Sec.~V A, the combination of $s$\,-wave and $p_y$-wave Cooper pairs carry the spontaneous
current in both chiral $p$\,- and chiral $d$\,-wave disks.  Therefore, the robust
spontaneous edge current and the robust spontaneous magnetization are common features in these two superconductors.
In the case of a chiral $f$-wave superconductor, the amplitude of the magnetization is almost
zero in the scale of Fig.~\ref{fig:chi_rgh}.  Within the numerical accuracy, we estimate that
the magnetization is smaller than $4\times 10^{-5} H_{c_2}$.


\section{Conclusion}

We have studied the effects of surface roughness on the spontaneous edge current in small
chiral superconductors characterized by chiral $p$\,-, chiral $d$\,- and chiral $f$-wave
pairing symmetries.  On the basis of the quasiclassical Eilenberger formalism, we calculated the
chiral current and the spontaneous magnetization of the small superconducting disk numerically.
By solving the Eilenberger and Maxwell equations simultaneously, we obtained 
self-consistent solutions of the pair potential, the impurity self-energy, and the vector
potential.  To understand the physics behind the complicated current distribution in real
space, we decomposed the current into a series of components in terms of the symmetry of a
Cooper pair.  The chiral edge current is carried by a combination of two pairing components: the
even-parity component and the odd-parity component.  In a spin-singlet (spin-triplet)
superconductor, the odd-parity (even-parity) Cooper pairs have odd-frequency symmetry.

The effects of the surface roughness depend on the pairing symmetry of the superconductor.  The
chiral current is robust in the presence of surface roughness in a chiral $p$\,- and chiral
$d$\,-wave symmetries. With chiral $p$\,-wave symmetry, the characteristic features of the
chiral current are insensitive to the surface roughness.  With chiral $d$\,-wave symmetry, the
chiral current changes its direction as a result of the surface roughness.  In both the chiral
$p$\,-wave and chiral $d$\,-wave cases, the chiral current is carried by a combination
consisting of two pairing correlations. One is the correlation with $p$\,-wave symmetry and the
other is the correlation with $s$\,-wave symmetry.  The Meissner screening effect by the bulk
condensate reduces {a} spontaneous magnetization. However, the resulting amplitude of the
magnetization is still large enough to be detected in experiments.  In a chiral $f$-wave
superconductor, the surface roughness significantly suppresses the spontaneous edge current.


\begin{acknowledgments}
The authors are grateful to Y.~Maeno, A.~A.~Golubov, Y.~Tanaka, S.~Kashiwaya, M.~Ichioka, Ya.~V.~Fominov, and 
M.~Yu.~Kupriyanov for
useful discussions. This work was supported by “Topological Materials Science” (No. 15H05852)
and KAKENHI (Nos. 26287069 and 15H03525) from the Ministry of Education, Culture, Sports,
Science and Technology (MEXT) of Japan, and by the Ministry of Education and Science of the
Russian Federation (Grant No. 14Y.26.31.0007).
S.-I.~S. is supported in part by Grant-in-Aid for JSPS Fellows (Grant No. 15J00797) by Japan
Society for the Promotion of Science (JSPS).
\end{acknowledgments}


%
%
%
%
%
%
%
%
%
%

\begin{thebibliography}{99}
\bibitem{Comment}
In this paper, we assume only the principal pairings (e.g., $p_x$- and $p_y$-wave pairs in a
chiral $p$\,-wave superconductor) are attractive.  Nevertheless, subdominant pairing
correlations belonging to different symmetry classes from that of the dominant one appear near
a surface.  The pairing correlations $f$ and the order parameters $\Delta$ are directly
connected by the gap equation Eq.~(\ref{gap}).  Subdominant order parameters never appear in our
formulation because we set the attractive interactions to be
zero for subdominant pairing correlations. These subdominant pairing correlations
induced by a surface are not related to the stability of a superconductor
because they can not become order parameters.
\end{thebibliography}
\end{document}